# Orbital two-channel Kondo effect in epitaxial ferromagnetic $L1_0$-MnAl films

L. J. Zhu,[1,2] S. H. Nie,[1] P. Xiong,[3] P. Schlottmann,[3] J. H. Zhao*[1]

[1]*State Key Laboratory of Superlattices and Microstructures, Institute of Semiconductors,*
*Chinese Academy of Sciences, P. O. Box 912, Beijing 100083, China*
[2]*Institut für Physik, Martin-Luther-Universität Halle-Wittenberg, von-Danckelmann-Platz 3, Halle 06120, Germany*
[3]*Department of Physics, Florida State University, Tallahassee, Florida 32306, USA*

**The Kondo effect is a striking consequence of the coupling of itinerant electrons to a localized quantum spin impurity with a degenerate ground state. In the intricate scenario of two conduction electrons compensating a spin-1/2 impurity, the overscreened two-channel Kondo (2CK) effect displaying exotic non-Fermi liquid behavior arises. As direct observation of the spin 2CK effect is difficult due to the strict requirements of zero local magnetic field and channel symmetry, a different type of 2CK effect with a pseudo-spin-1/2 impurity of two-level system (TLS) was proposed, which is termed orbital 2CK effect. Despite extensive experimental efforts for several decades, no material system has been clearly identified to exhibit all three transport temperature regimes characteristic of the 2CK effect, making the interpretation of the experimental results remain a subject of debate. Here we present an experimental realization of highly robust orbital 2CK effect in epitaxial ferromagnetic $L1_0$-MnAl films, as evidenced by a low-temperature ($T$) resistivity upturn with a clear transition from a ln$T$-dependence to $T^{1/2}$-dependence and deviation from it in three distinct $T$ regimes, which are independent of applied magnetic fields up to 8 T. The ability to tune the density of the TLSs, Kondo temperature ($T_K$) and energy splitting of the TLSs ($\Delta$) in this material system by varying growth parameters leads to new insights on the origin of the orbital 2CK effect. The $T_K$ and $\Delta$ are greatly enhanced in comparison to those in other systems exhibiting orbital 2CK, suggesting strong coupling between the tunneling centers with conduction electrons via resonant scattering. Our results also provide the first evidence for the presence of 2CK physics in a ferromagnet, pointing to considerable robustness of the orbital 2CK effect even in the presence of ferromagnetic ordering and spin polarization of the conduction electrons.**

The overscreened two-channel Kondo (2CK) effect displaying exotic non-Fermi-liquid (NFL) physics has attracted widespread interest in recent years, especially due to their potential relevance to a host of current topics including strongly correlated physics, Majorana fermions, high-$T_c$ superconductors, topological insulators, and quantum computation[1-6]. In spin 2CK effect, a spin-1/2 impurity couples to conduction electrons in two equal orbital channels via an exchange interaction[6-9]. Below Kondo temperature ($T_K$) the 2CK model gives rise to impurity quantum criticality accompanied by exotic NFL behavior as the consequence of *two* conduction electron spins attempting to compensate the spin-1/2 impurity. However, the strict requirements of zero local magnetic field and channel symmetry make a direct observation of the *spin* 2CK effect difficult. Intriguingly, an analogous *orbital* 2CK effect was proposed to arise from scattering centers with orbital degrees of freedom, e.g. two-level systems (TLSs)[10]. As depicted in Fig. 1a, in a TLS, the tunneling entity (e.g., an atom, atom group, or localized electron) coherently tunnels at a rate of $10^8 \sim 10^{12}$ s$^{-1}$ between two independent quantum wells with asymmetry energy $\Delta_z$, tunneling matrix element $\Delta_x$, and energy splitting $\Delta=(\Delta_z^2+\Delta_x^2)^{1/2}$ between the lowest two eigenstates[11,12]. For the *noncommutative* model this problem is reduced to the 2CK case, the TLS being represented by a pseudospin 1/2 and the spin degeneracy of the conduction electrons being the two channels[10,13]. The orbital 2CK effect from TLSs is manifested in electrical transport by a unique temperature ($T$)-dependence in resistivity with three distinct $T$ regimes (Fig. 1b): a low-$T$ upturn characterized by $\Delta\rho \sim \ln T$ for $T > T_K$, followed by NFL behavior ($\Delta\rho \sim T^{1/2}$) for $T_D$ ($=\Delta^2/T_K$) $< T \ll T_K$ and deviation from $T^{1/2}$ dependence upon further cooling[11,12]. The $T^{1/2}$ dependence is a hallmark of the NFL state in the orbital 2CK effect, in striking contrast to the $T^2$ scaling of Fermi-liquid (FL) behavior in fully screened Kondo effect[14]. The stability of the low $T$ orbital 2CK fixed point is an important issue. Theoretically, the fixed point is unstable to $\Delta$, the strength of Kondo coupling between TLSs and itinerant electrons, channel-symmetry breaking in the exchange coupling, and a magnetic field changing the electron population of the two channels, but stable to exchange anisotropy in the Kondo coupling[15-18].

Despite the intensive studies for almost 30 years, the experimental proof for the existence of the orbital 2CK effect has been far from certain. First, it remains a challenge to unambiguously demonstrate the three-regime $T$-dependence of resistivity of the orbital 2CK effect in a single material system. Although the NFL behavior was reported in the amorphous nonmagnetic Cu point contacts (PCs) and single-crystalline diamagnetic ThAsSe glasses[19,20], their small $\Delta$ prevented the observation of possible breakdown of the $T^{1/2}$ behavior at lower $T$. Upadhyay et al.[21] observed the NFL behavior and the low-energy restoration of the FL in conductance spectroscopy of Ti point contacts, but there was no indication of a crossover to logarithmic dependence at high energies. Furthermore, the $H$-dependence and microscopic mechanisms of the orbital 2CK effect have remained unsettled. Significant negative magneto-resistance (MR) and $H$-suppression of $T_K$ were also reported in Cu PCs[19]. In ThAsSe, a magnetic field of up to 14 T did not affect the NFL behavior[20], although present theories expect an imbalance in the channel population to produce a crossover to FL behavior[22]. Microscopically, the dynamic tunneling centers in TLSs were interpreted as a group of nonmagnetic atoms in PCs[19], whereas resonant tunneling of an electron in a polar bond transformation was argued to be responsible for the enhancement of $T_K$ to a few K in the metallic glasses[20]. Therefore, materials with TLSs of larger $T_K$ and $\Delta$ are desirable for a thorough study of the orbital 2CK physics, including its $T$ and $H$ dependencies and microscopic mechanisms.

Ferromagnetic correlations are predicted to have a detrimental effect on the spin 2CK, while in the case of the orbital 2CK, a coexistence over a large $T$-range is possible as the electron spin variable does not directly enter into the interaction process. A crossover to FL behavior is, however, expected at low temperatures due to the asymmetry in the channel population. So far, there are two reports on orbital Kondo effect in ferromagnets[23,24]. The ferromagnetic UAs$_{1-x}$Se$_{1+x}$ and Mn$_5$Si$_3$C$_x$ display a logarithmic low-$T$ upturn in the resistivity. A crossover to FL behavior was also observed in Mn$_5$Si$_3$C$_x$ at low $T$. However, neither shows a $T^{1/2}$-dependence characteristic of 2CK effect, possibly because of their low $T_K$, large $T_D$ and partial polarization of the conduction electrons.

In this Letter, we report the first experimental evidence of TLS-induced orbital 2CK effect in a ferromagnetic system, $L1_0$-MnAl epitaxial films with strong perpendicular magnetic anisotropy (PMA). We observed a low-$T$ resistivity upturn with a clear transition from a ln$T$-dependence to NFL behavior signified by a $T^{1/2}$-dependence, and deviation from it upon further cooling. The $T$-dependencies are independent of applied magnetic fields up to 8 T. This represents the first observation of all three theoretically-expected transport regimes from the orbital 2CK effect in the same samples. The greatly enhanced $T_K$ and $\Delta$ in this system suggest fast coherent tunneling of the TLSs and strong coupling with the conduction electrons. Moreover, the structural disorders in $L1_0$–ordered films can be tailored by varying the growth parameters[25-29], offering a convenient pathway of tuning the relevant 2CK parameters.

A series of 30 nm-thick $L1_0$-MnAl single-crystalline films (Fig. 1c) were grown at substrate temperatures ($T_s$) of 200, 250, 300, 350 and 400 °C[30]. The degree of structural disorder decreases with increasing $T_s$ from 200 to 350 °C, and then increases when $T_s$ goes up to 400 °C, which is evidenced by the full width at half maximum of the MnAl (002) peaks of x-ray diffraction patterns[30]. These films exhibit strong PMA as revealed by the well-defined hysteretic anomalous Hall resistance (Fig. 1d) and perpendicular magnetization hysteresis[31].

Figure 1e shows the $T$-dependence of the zero-field longitudinal resistivity ($\rho_{xx}$) of the $L1_0$-MnAl films. Each film shows a resistivity minimum at a characteristic temperature ($T_m$). In the high $T$ regime ($T > T_m$), $\rho_{xx}$ increases linearly with $T$ due to increasing phonon scattering[30]. Here we show that the low-$T$ resistivity upturn in our $L1_0$-MnAl films likely arises from the TLS-induced orbital 2CK effect. Figures 2a-2c plot the $T$-dependence of resistivity variation at $H$=0 T for the $L1_0$-MnAl films, which shows distinct signatures associated with the TLS-induced 2CK effect. In the first regime, as displayed in Fig. 2a, the resistivity increase, $\Delta\rho_{xx}$ ($\Delta\rho_{xx}=\rho_{xx}-\rho_1$, with the offset $\rho_1$ determined from the best linear fit of $\rho_{xx}$-ln$T$)[30], varies linearly with ln$T$ below a temperature $T_0$ for all films with different $T_s$, similar to the well-known single-channel Kondo (1CK) effect due to static magnetic impurities[14]. $\Delta\rho_{xx}$ deviates from the ln$T$ dependence and transitions to a $T^{1/2}$ dependence when $T$ drops below $T_K$. The $T^{1/2}$-dependent resistivity is regarded as a distinct signature of the NFL behavior from the 2CK effect.

As can been seen in Fig. 2b, $\Delta\rho_{xx}$ ($\Delta\rho_{xx}=\rho_{xx}-\rho_2$, where $\rho_2$ was determined from the best linear fit of $\rho_{xx}$-$T^{1/2}$; $\rho_1$ and $\rho_2$ track each other[30]) begins to increase more slowly than $T^{1/2}$ below a characteristic temperature $T_D$, indicating deviation from the NFL behavior. The deviation is generic, $H$-independent (Fig. 3a), and distinct from that induced by quantum corrections in ThAsSe which only appeared at zero field[20]. This is, to the best of our knowledge, the first observation of the TLS theory-expected deviation from the orbital 2CK state below $T_D$ in a diffusive conductor. We emphasize that the deviation is not due to Joule heating because the resistivities measured under AC current of 1 $\mu$A and DC current of 10 $\mu$A are virtually identical. The lower temperature limit for the $T^{1/2}$ regime is given theoretically by $T_D=\Delta^2/T_K$, from which $\Delta$ can be determined. In fully screened 1CK systems[14] and "frozen" slow two-state systems[24,32], $\rho_{xx}$ was observed to saturate following the FL behavior (~$T^2$) at low $T$. The latter have large tunneling barrier and negligibly small $\Delta_x$ that only allow thermally-activated hopping or incoherent tunneling at a very slow rate (<<$10^8$ s$^{-1}$)[10]. However, in the TLS-induced 2CK effect, it has remained unclear how the system experimentally deviates from the NFL behavior below $T_D$. In order to shed light on this important issue in $L1_0$-MnAl, we measured $\rho_{xx}$ down to 330 mK. As shown in Fig. 2c, none of these films with different $T_s$ shows FL-like saturation ($T^2$) in the studied temperature range, distinctly different from the expectation of a fully screened Kondo effect[14]. There are three possible reasons for this discrepancy: (1) a broad distribution of $T_D$ of the TLSs, (2) overlapping of the screening clouds of different TLSs, and (3) the spin polarization of the conduction electrons due to the ferromagnetism. A broad distribution of $T_D$ would severely distort the $T^{1/2}$-behavior and is not a likely scenario. The 2CK displays impurity quantum critical behavior so that the screening would diverge at $T = 0$ under ideal conditions ($\Delta = 0$), in contrast to the 1CK problem where the screening radius ($R$) is finite. For orbital 2CK with nonzero $\Delta$, $R$ at $T =$ 0 K limit can be estimated by $k_FR\sim D/T_D$, where $k_F$ is Fermi wavevector, $D$ is band width (of the order of Fermi Energy $E_F$). Using $k_F \sim 1.7$ Å$^{-1}$ ($E_F \sim 11$ eV)[33], $D \sim 10$ eV, $T_D \sim 1$ K for MnAl, one can estimate $R$ to be on the order of $\sim 10^4$ Å at $T = 0$ K, which is much larger than the average distance between TLSs (see below). Hence, with decreasing $T$ the screening clouds will eventually overlap, although this problem has yet to be quantitatively studied. The spin-polarization, if homogeneous, should produce a crossover to a $T^2$-dependence.

Figure 2d plots the slopes $\alpha = -d\rho_{xx}/d(\ln T)$ for $T_K < T < T_0$ and $\beta = -d\rho_{xx}/d(T^{1/2})$ for $T_D < T < T_K$ as a function of $T_s$. It is evident that $\alpha$ and $\beta$ have a similar $T_s$ dependence, i.e., $\alpha$ ($\beta$) first drops quickly from their maximum of 2.5 μΩ cm/lnK (1.0 μΩ cm/K$^{1/2}$) at 200 °C to the minimum of 0.2 μΩ cm/lnK (0.15 μΩ cm/K$^{1/2}$) at 300~350 °C, and finally goes up to 0.25 μΩ cm/lnK (0.18 μΩ cm/K$^{1/2}$) at 400 °C. As is revealed in the metallic PC experiments, thermal annealing can significantly change the number of TLSs[19]. Here, the non-monotonic dependence of $\alpha$ and $\beta$ on $T_s$ could be ascribed mainly to the thermal-tailoring of the density of active TLSs ($N_{TLS}$). For the strong coupling TLS centers[11], $N_{TLS}$ can be estimated by $N_{TLS} \sim \frac{\Delta \rho_{xxm}}{\rho_{xx}} \frac{N(E_F)}{\tau_e}$ in the diffusive transport regime, where $\Delta \rho_{xxm}$, $\tau_e$ and $N(E_F)$ are the maximum resistivity upturn due to the TLSs, electron scattering time and density of states at Fermi level, respectively. $\tau_e$ can be determined to be $\sim 10^{-15}$ s by $\rho_{xx}=m^*/ne^2\tau_e$, where $m^*$, $n$, and $e$ are effective mass, density ($\sim 10^{22}$ cm$^{-3}$) and charge of conduction electrons, respectively. Using a typical $N(E_F)$[33] of $\sim 4 \times 10^{22}$ eV$^{-1}$cm$^{-3}$ and experimental values of $\Delta\rho_{xxm}$ ($\approx \Delta\rho_{xx}$ at 330mK), $N_{TLS}$ was calculated and shown in Fig. 1d. The fast TLSs were annealed away[19] and its density was reduced quickly as $T_s$ increases to 300 °C; however, the overall population of TLSs increases beyond 350 °C due to the structural deterioration and strain relaxation of the $L1_0$-MnAl films[30]. The large $N_{TLS}$ ($\sim 10^{20}$ cm$^{-3}$) yields an average distance of $\sim 20$ Å for the TLSs in these films.

In order to establish more rigorously the orbital 2CK effect in our $L1_0$-MnAl films, we examined the effect of applied perpendicular magnetic fields, $H$, on the $T$-dependent resistivity. Here, in these films with strong PMA, anisotropic MR and MR from spin disorder scattering under perpendicular $H$ should be negligible due to the orthogonal magnetization-current relation and the large energy gap in spin wave excitation spectrum. This is highly amenable to study the $H$-dependence of a 2CK effect. As an example, we show $\rho_{xx}$ ($T$) of the $L1_0$-MnAl film grown at 200 °C under various $H$ from 0 to 8 T in Figs. 3a and 3b. The magnetic fields have no measurable effects on the $T$-dependence: $\rho_{xx}$ scales linearly with $\ln T$ and $T^{1/2}$ at $T_K<T<T_0$ and $T_D<T<T_K$ ($T_0 \sim 82.5$ K, $T_K \sim 23$ K, and $T_D \sim 8.8$ K), respectively. The same features hold for other films with different $T_s$. Figures 3c and 3d summarize the values of $\alpha$ and $\beta$ as a function of $H$ for the $L1_0$-MnAl films with different $T_s$. It is clear that both $\alpha$ and $\beta$ for all films are independent of $H$, strongly suggesting a nonmagnetic origin of the resistivity upturn scaling in the $L1_0$-MnAl. Specifically, there is no measurable change in $T_0$, $T_K$ and $T_D$ under different $H$ (Figs. 3a and 3b), suggesting a negligible effect of $H$ on the tunneling symmetry and barrier height of the TLSs. These observations provide strong evidence for the orbital 2CK effect being induced by TLSs of nonmagnetic impurities. Here, it also should be pointed out that the Zeeman energy ($\sim 5$ meV at $H=8$ T) is negligibly small in comparison to $E_F$ and ferromagnetic exchange splitting ($E_{exchange} \sim 2$eV) in $L1_0$-MnAl[33], hence should not have any measurable effect on the channel asymmetry ($\Delta N = N_\uparrow - N_\downarrow$, where $N_\uparrow$ and $N_\downarrow$ are the numbers of majority and minority spins in the conduction band, respectively) and the three-regime resistivity upturn[16,21,22]. A small negative MR (<0.5%) is observed in the $L1_0$-MnAl at high $H$ in the entire $T$ range[31]. The negative MR shows a $T$-dependence which appears to have no correlation with any characteristic temperatures of the 2CK effect (Fig. 4). The negative MR does not saturate even at 7 T and shows a linear scaling with $H^{1/2}$ instead of $H^2$, which is similar to that in metallic PCs[12], but in contrast to that in metallic glasses[20]. Though not yet well understood, such a diverse MR behavior in materials showing orbital 2CK effect due to TLSs may be additional evidence of the nonmagnetic origin.

Figure 4 summarizes the relevant characteristic temperatures $T_0$, $T_K$, $T_D$, and $\Delta$ of the $L1_0$-MnAl films as a function of $T_s$. As $T_s$ increases, $T_0$ drops dramatically from 82.5 K to 11.4 K, and goes up to 20.0 K as a consequence of the non-monotonic dependence of the TLS population on $T_s$ (Fig. 2d). $T_K$ represents the temperature below which conduction electrons can overscreen the "pseudo-spin" of the impurity in an orbital 2CK system. Since there is an overlap between the two $T$-regimes of $\ln T$ and $T^{1/2}$ dependencies, the values of $T_K$ are defined as the center of the $T$-overlap. In striking contrast to the small experimental values in Cu PCs (0.1~5 K) and theoretical calculations on amorphous systems (<<1 K), $T_K$ ranges from 23.0 K to 5.4 K in the $L1_0$-MnAl. The significant enhancement of $T_K$ in this system could be understood in terms of resonant scattering due to strong coupling of conduction electrons to the scattering centers[18]. Within this scenario, the decrease of $T_K$ at higher $T_s$ may come from a reduced strength of the resonant scattering. Another intriguing observation is that $\Delta$ is tuned by as much as a factor of ~5.6 when $T_s$ varies between 200 and 400 °C, which is difficult to achieve in previously studied systems of metallic PCs or glasses using a voltage or magnetic field[19-21]. As $T_s$ goes up, the fastest TLSs are annealed away, leaving the slower ones to dominate the scattering of conduction electrons. Therefore, with $T_s$ increasing from 200 °C to 350 °C, the value of $\Delta$ for active TLSs decreases, leading to a reduction of $T_D$. The upturn of $T_0$, $T_K$, $T_D$, and $\Delta$ at $T_s$ of 400 °C may be attributed to the increased population of fast TLSs and enhanced resonant scattering of conduction electrons due to the structural degradation[30]. Here, it is worth mentioning that the values of $\Delta$ observed in $L1_0$-MnAl films are much larger than the reported values in other 2CK systems, e.g. $\Delta \sim 1$ K in PCs of quenched Cu[19] and $\Delta < 0.01$ K in ThAsSe[20]. The disorders in amorphous or polycrystalline systems tend to produce TLSs with small $\Delta$[19] or even slow two-state systems with incoherent tunneling[24,32]. This may be why the NFL behavior was not observed in amorphous or polycrystalline systems. In contrast, in the MBE-grown epitaxial $L1_0$-MnAl films, the greatly enhanced $\Delta$ indicates the formation of TLSs with small tunnel barriers and very fast scattering rates, which further leads to the enhanced resonant tunneling and high $T_K$. As suggested by the close correlation between the $T_s$ dependence of the characteristic parameters of the TLS-induced 2CK effect, i.e. $α$, $β$, $T_0$, $T_K$, and $Δ$ with those of structural imperfections[30] and the magnetic properties (e.g. magnetization and PMA)[31], electrons as the tunneling centers of the TLSs in the $L1_0$-MnAl films can be excluded. Similarly, the magnetic Mn atoms can also be ruled out because of the nonmagnetic nature of the orbital Kondo effect. Therefore, we surmise that the nonmagnetic small Al atoms play the role of TLSs centers in the present films.

Now we discuss the apparent coexistence of orbital 2CK with ferromagnetism. In a conventional ferromagnet, although the Kondo coupling between the TLS and itinerant electrons is irrelevant to the electron spins is still independent of the electronic spin, the symmetry of the two spin channels is broken due to ferromagnetic exchange splitting of the $d$-band. The channel asymmetry should lead to different tunneling rates of a TLS for two spin channels and thus weaken the NFL behaviors in comparison to its nonmagnetic counterpart ($ΔN=0$). If the degree of channel asymmetry ($P=ΔN/(N_↑+N_↓)$) is large enough, this should be manifested as a decreased magnitude ($β$) and an enhanced effective breakdown temperature ($T_D$) of the 2CK effect. The latter is due to the competition between the 2CK physics and Kondo cloud overlap at low temperatures. Based on a simple assumption in the Stoner model for itinerant ferromagnetism[34], the saturation magnetization ($M_s$) can be an index of the channel asymmetry, i.e. $M_s ∝ ΔN$. However, this effect is difficult to quantify experimentally as one cannot vary $M_s$ while keeping other parameters (e.g. the TLS density, $T_K$ and $Δ$) constant in a set of samples. For our $L1_0$-MnAl films, the $T_s$-dependence of the measured $β$ and $T_D$ is dominated by the *intrinsic* variation of the density and $Δ^2/T_K$ of the TLSs, rather than that of $ΔN$, as suggested by the quite different $T_s$-dependence of the 2CK parameters ($β$ and $T_D$) and $M_s$. The fully ordered $L1_0$-MnAl is theoretically an itinerant magnet with a magnetization $M_s$ of ~800 emu cm$^{-3}$ (i.e. 2.37 $μ_B$ f.u.$^{-1}$)[25,33]. In our disordered MnAl samples, the measured $M_s$ is much smaller (157 - 306 emu cm$^{-3}$, see Fig. 5), suggesting a very low $ΔN$ ($<<N_↓$) and $P$ ($<<1$). This could be a reason why it is still difficult to establish a definitive quantitative correlation between $ΔN$ and the 2CK parameters ($β$ and $T_D$).

In conclusion, we have for the first time observed transport behavior suggestive of a robust orbital 2CK effect from TLSs in epitaxial $L1_0$-MnAl films with strong PMA. The $H$-independent resistivity upturn scaling with $\ln T$ and $T^{1/2}$ in the two $T$ regimes below the resistance minimum, and deviation from the NFL behavior at the lowest temperatures are consistent with the TLS model. The MBE growth method afforded unprecedented tunability of the TLS density, $T_K$ and $Δ$ in this material system, leading to the new insight on the origin of the 2CK effect. The greatly enhanced $T_K$ and $Δ$ suggest resonant tunneling of TLSs due to the strong coupling with conduction electrons. The orbital 2CK effect in a material with strong ferromagnetism and significant conduction spin polarization is an intriguing observation that warrants further theoretical and experimental studies.

## Methods

**Sample growth and characterization.** $L1_0$-MnAl films were grown on 150 nm GaAs-buffered semi-insulating GaAs (001) by molecular-beam epitaxy and capped with a~4 nm-thick Al$_2$O$_3$ layer[30]. The thickness and Mn/Al atomic ratio of MnAl layer is calibrated to be 30 nm and 1.1 by cross-sectional high-resolution transmission microscopy with EDX[28], respectively. Synchrotron x-ray diffraction and Quantum Design SQUID were used to characterize the structures and magnetization properties of these films (see Supplementary Fig. S1).

**Device fabrication and transport measurement.** These films were patterned into 60 μm wide Hall bars with adjacent electrode distance of 200 μm using photolithography and ion-beam etching for transport measurements (Fig. 1c). The longitudinal resistivity was measured as a function of temperature ($T$) and magnetic field ($H$) in a Quantum Design PPMS for $T$ = 2-300 K (DC, $I$ =10 $μA$) and in an Oxford $^3$He cryostat for $T$ = 0.33-5 K (AC, $I$ =1 $μA$).

## References

1  Béri, B. & Cooper, N. R. Topological Kondo effect with Majorana Fermions. *Phys. Rev. Lett.* **109**, 156803(2012).
2  Kashuba, O. & Timm, C. Topological Kondo effect in transport through a superconducting wire with multiple Majorana end states, *Phys. Rev. Lett.* **114**, 116801 (2015).
3  Oreg Y. & Goldhaber-Gordon, D. Two-channel Kondo effect in a modified single electron transistor. *Phys. Rev. Lett.* **90**, 136602 (2003).
4  Mitchell, A. K., Sela, E. & Logan, D. E. Two-channel Kondo physics in two-Impurity Kondo models. *Phys. Rev. Lett.* **108**, 086405 (2012).
5  Cox, D. L. Quadrupolar Kondo effect in uranium heavy-electron materials. *Phys. Rev. Lett.* **59**, 1240 (1987).
6  Potok, R. M., Rau, I. G., Shtrikman, H., Oreg, Y. & Goldhaber-Gordon, D. Observation of the two-channel Kondo effect. *Nature* **446**, 167 (2007).
7  Yeh, S. & Lin, J. Two-channel Kondo effects in Al/AlO$_x$/Sc planar tunnel junctions. *Phys. Rev. B* **79**, 012411 (2009).
8  Nozières, P. & Blandin, A. Kondo effect in real metals. *J. Phys. France* **41**, 193-211(1980).
9  Schlottmann, P. & Sacramento, P. D. Multichannel Kondo problem and some applications. *Adv. Phys.* **42**, 641 (1993).
10  Zawadowski, A. Kondo-like state in a simple model for metallic glasses. *Phys. Rev. Lett.* **45**, 211 (1980).
11  Cox, D. L. & Zawadowki, A. Exotic Kondo effects in metals: Magnetic ions in a crystalline electric field and tunneling centres. *Adv. Phys.* **47**, 599 (1998).
12  Delft, J. von, *et al.*, The 2-channel Kondo model: I. Review of experimental evidence for its realization in metal nanoconstrictions. *Annals. Phys.* **263**, 1 (1998).
13  Muramatsu, A. & Guinea, F. Low-temperature behavior of a tunneling atom interacting with a degenerate electron gas. *Phys. Rev. Lett.* **57**, 2337 (1986).
14  Hewson, A. C. *The Kondo Problem to Heavy Fermions* (Cambridge University Press, Cambridge, 1993).
15  Pang, H. B. & Cox, D. L. Stability of the fixed point of the two-channel Kondo Hamiltonian. *Phys. Rev. B* **44**, 9454 (1991).
16  Affleck, I. *et al.*, Relevance of anisotropy in the multichannel Kondo effect: Comparison of conformal field theory and numerical renormalization-group results. *Phys. Rev. B* **45**, 7918 (1992).
17  Aleiner, I. L. *et al.*, Kondo temperature for the two-channel Kondo models of tunneling centers. *Phys. Rev. Lett.* **86**, 2629 (2001); Experimental tests for the relevance of two-level systems for electron dephasing, *Phys. Rev. B* **63**, 201401 (R) (2001).

18 Zaránd, G., Existence of a two-channel Kondo regime for tunneling impurities with resonant scattering. *Phys. Rev. B* **72**, 245103 (2005).

19 Ralph, D. C. & Buhrman, R. A. Observations of Kondo scattering without magnetic impurities: A point contact study of two-level tunneling systems in metals. *Phys. Rev. Lett.* **69**, 2118 (1992); Ralph, D. C., Ludwig, A. W. W., Delft, J. v. & Buhrman, R. A. 2-channel Kondo scaling in conductance signals from 2 level tunneling systems. *Phys. Rev. Lett.* **72**, 1064 (1994).

20 Cichorek, T., et al., Two-channel Kondo effect in glasslike ThAsSe. *Phys. Rev. Lett.* **94**, 236603 (2005); Cichorek, T., et al., TLS Kondo effect in structurally disordered ThAsSe. *J. Mag. Mag. Mater.* **272**, 67 (2004).

21 Upadhyay, S. H., et al., Low-energy restoration of Fermi-liquid behavior for two-channel Kondo scattering. *Phys. Rev. B* **56**, 12033 (1997).

22 Schlottmann, P. & Lee, K. J. B. Quenching of overcompensated Kondo impurities via channel asymmetry. *Physica B* **223&224**, 458 (1996).

23 Henkie, Z., Fabrowski, R. & Wojakowski, A. Anisotropies of the electrical resistivity and Hall effect in UAsSe. *J. Alloys and Compounds* **219**, 248 (1995).

24 Gopalakrishnan, B., Sürgers, C., Montbrun, R., Singh, A., Uhlarz, M. & Löhneysen, H. v. Electronic transport in magnetically ordered $Mn_5Si_3C_x$ films. *Phys. Rev. B* **77**, 104414 (2008).

25 Zhu, L. J., Nie, S. H. & Zhao, J. H. Recent progress in perpendicularly magnetized Mn-based binary alloy films. *Chin. Phys. B* **22**, 118505 (2013).

26 Zhu, L. J., Nie, S. H., Meng, K. K., Pan, D., Zhao, J. H., & Zheng, H. Z., Multifunctional $L1_0$-$Mn_{1.5}Ga$ films with ultrahigh coercivity, giant perpendicular magnetocrystalline anisotropy and large magnetic energy product. *Adv. Mater.* **24**, 4547 (2012).

27 Zhu, L. J., Pan, D., Nie, S. H., Lu, J., & Zhao, J. H. Tailoring magnetism of multifunctional $Mn_xGa$ films with giant perpendicular anisotropy. *Appl. Phys. Lett.* **102**, 132403 (2013).

28 Nie, S. H., Zhu, L. J., Lu, J., Pan, D., Wang, H. L., Yu, X. Z., Xiao, J. X., & Zhao, J. H. Perpendicularly magnetized τ-MnAl (001) thin films epitaxied on GaAs. *Appl. Phys. Lett.* **102**, 152405 (2013).

29 Zhu, L. J., Pan, D., & Zhao, J. H. Anomalous Hall effect in epitaxial $L1_0$-$Mn_{1.5}Ga$ films with variable chemical ordering. *Phys. Rev. B* **89**, 220406 (R) (2014).

30 See Supplementary Materials at xxxxxx for further details on structural, magnetic, electrical transport and magnetoresistance behaviors of $L1_0$-MnAl films.

31 Nie, S. H., Zhu, L. J., Pan, D., Lu, J., & Zhao, J. H. Structural characterization and magnetic properties of perpendicularly magnetized MnAl films grown by molecular-beam epitaxy. *Acta. Phys. Sin.* **62**, 178103 (2013) (in Chinese).

32 Zimmerman, N. M., Golding, B. & Haemmerle, W. H. Magnetic field tuned energy of a single two-level system in a mesoscopic metal. *Phys. Rev. Lett.* **67**, 1322 (1991).

33 Sukuma, A., Electronic structure and magnetocrystalline anisotropy Energy of MnAl, *J. Phys. Soc. Jpn.* **63**, 1422 (1994).

34 Stoner, E. C., Collective electron ferromagnetism. Proc. R. Soc. London, Ser. A **165**, 372 (1938).

**Acknowledgements**

We thank Y. Q. Li and J. Liao for their help on the ultralow temperature measurements. We also thank S. von. Molnár and G. Woltersdorf for discussions. L.J.Z., S.H.N., and J.H.Z. were supported partly by MOST of China (Grant No. 2015CB921503), NSFC (Grant No. 61334006), and the CAS/SAFEA International Partnership Program for Creative Research Teams. P.X. acknowledges support from NSF grant DMR-1308613. P.S. was supported by the U.S. Department of Energy under grant DE-FG02-98ER45707.

**Author contributions**

L.J.Z and J.H.Z. designed the experiment. L.J.Z and S.H.N. grew the samples and performed the structural and magnetic characterizations. L.J.Z fabricated the Hall bar devices and performed transport measurements. L.J.Z, P.X., P.S. and J.H.Z analyzed the data and wrote the manuscript.

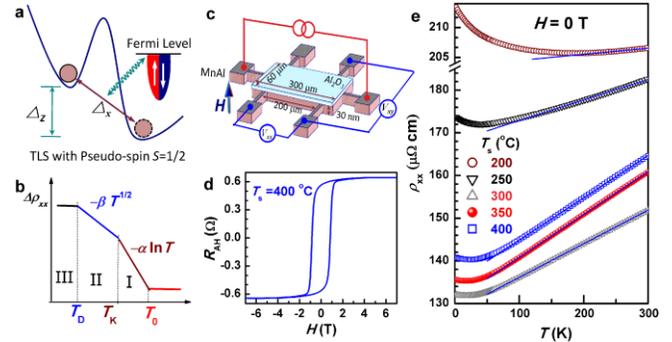

**Figure 1| Orbital 2CK effect from TLSs and typical transport properties of $L1_0$-MnAl films.** (**a**) Schematic depiction of a TLS with pseudo-spin $S=1/2$ and strong coupling with conduction electrons; (**b**) Expected $\rho_{xx}$-$T$ behavior from TLS-induced 2CK effect: $\Delta\rho_{xx}\sim\ln T$ for $T_K<T<T_0$ (I), $\Delta\rho_{xx}\sim T^{1/2}$ for $T_D (=\Delta^2/T_K) < T << T_K$ (II), and $T^2$-deviation from $\Delta\rho_{xx}\sim T^{1/2}$ for $T < T_D$ (III); (**c**) Schematics of the Hall bar device and measurement scheme; (**d**) Hysteretic anomalous Hall resistance ($R_{AH}$) ($T_s$=400 °C, $T$=300 K) and (**e**) $\rho_{xx}$-$T$ curves for $L1_0$-MnAl films.

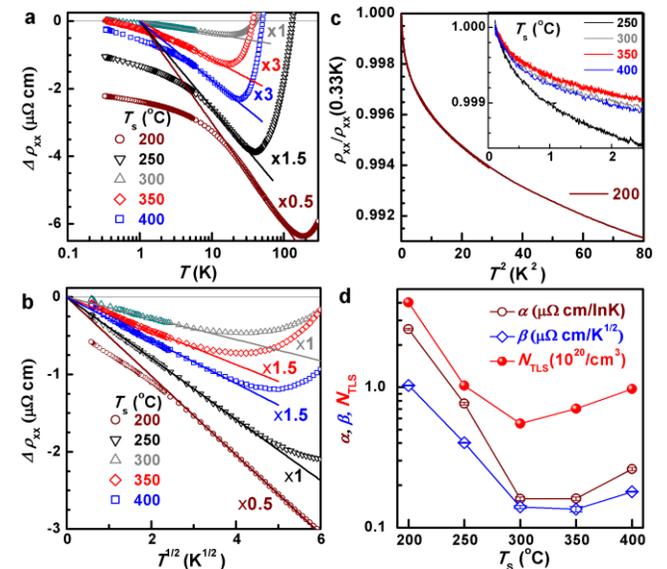

**Figure 2| $T$-dependence of zero field resistivity of the $L1_0$-MnAl films.** (**a**) Semilog plot of $\Delta\rho_{xx}$ versus $T$, (**b**) $\Delta\rho_{xx}$ versus $T^{1/2}$, (**c**) $\rho_{xx}/\rho_{xx}$(0.33 K) versus $T^2$, and (**d**) $\alpha$, $\beta$ and $N_{TLS}$ versus $T_s$. For clarity, $\Delta\rho_{xx}$ is multiplied by a factor of 0.5 (0.5), 1.5 (1), 1 (1), 3 (1.5), and 3 (1.5) in (**a**) ((**b**)) for $T_s$=200, 250, 300, 350 and 400 °C, respectively.

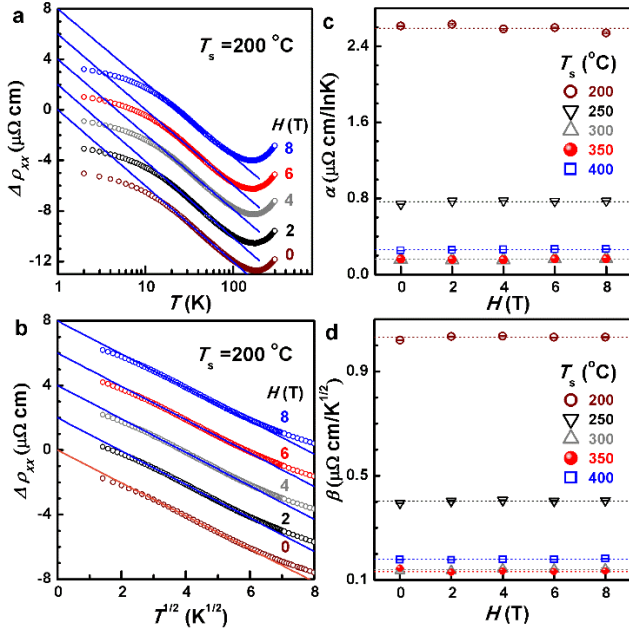

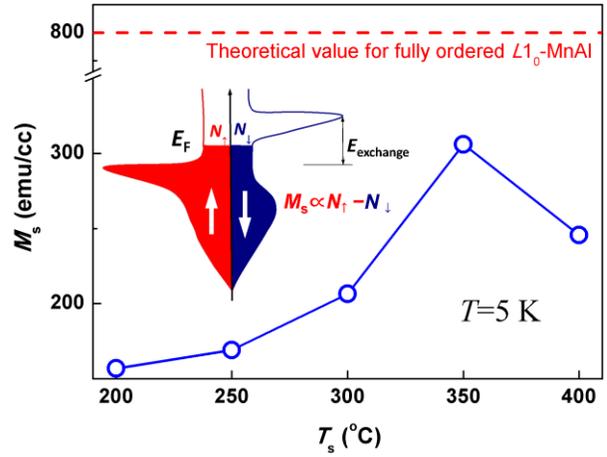

Figure 3| *T*-dependence of resistivity of the $L1_0$-MnAl films under different perpendicular magnetic fields. (a) Semilog plot of $\Delta\rho_{xx}$ versus *T* and (b) $\Delta\rho_{xx}$ versus $T^{1/2}$ for $T_s$=200 °C; (c) $\alpha$ and (d) $\beta$ plotted as a function of *H* for films with different $T_s$. For clarity, the curves in nonzero fields are artificially shifted by steps of 2 μΩ cm in (a) and (b).

Figure 5| $T_s$-dependence of saturation magnetization $M_s$ at 5 K in $L1_0$-MnAl films. The red dashed line shows the theoretical value of $M_s$ for fully ordered $L1_0$-MnAl films. The inset show a schematic of the partial density of state of a ferromagnet, where $E_F$, $E_{exchange}$, and $N_\uparrow$ ($N_\downarrow$) are the Fermi energy and the exchange splitting, and the number of the majority (minority) spin in conduction band, respectively.

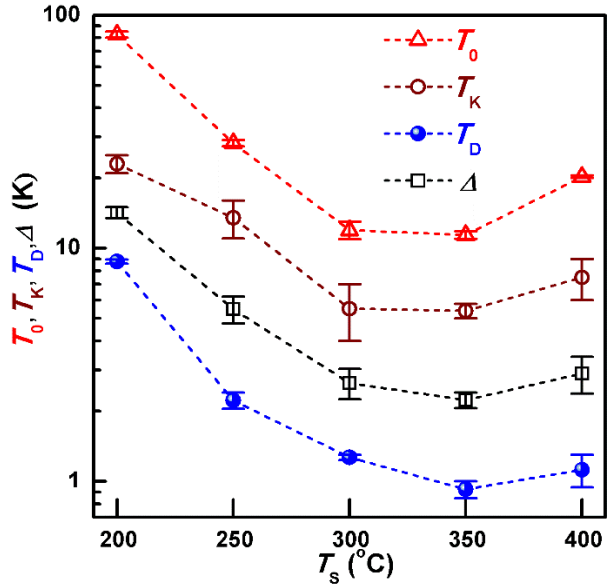

Figure 4| Characteristic temperatures $T_0$, $T_K$, $T_D$, and $\Delta$ in the $L1_0$-MnAl films plotted as a function of $T_s$.